\documentclass[preprint,aps, amsfonts, nofootinbib]{revtex4}

\usepackage{epsf}
\usepackage{graphicx}
\usepackage{amscd}
\usepackage{amsmath}
\usepackage{color}
\usepackage{comment}

\input xy
\xyoption{all}

\newcommand{\beq}{\begin{equation}}
\newcommand{\eeq}{\end{equation}}
\newcommand{\bea}{\begin{eqnarray}}
\newcommand{\eea}{\end{eqnarray}}

\newcommand{\nn}{\nonumber}
\newcommand{\benn}{\begin{displaymath}}
\newcommand{\eenn}{\end{displaymath}}

\def\slashchar#1{\ensuremath{                               %
   \setbox0=\hbox{${}#1{}$}       
   \dimen0=\wd0                                 
   \setbox1=\hbox{/} \dimen1=\wd1               
   \ifdim\dimen0>\dimen1                        
      \rlap{\hbox to \dimen0{\hfil/\hfil}}      
      {}#1{}                                    
   \else                                        
      \rlap{\hbox to \dimen1{\hfil${}#1{}$\hfil}}   
      /                                         
   \fi}}                                        %

\begin{document}
%
%
\newcount\hour \newcount\hourminute \newcount\minute 
\hour=\time \divide \hour by 60
\hourminute=\hour \multiply \hourminute by 60
\minute=\time \advance \minute by -\hourminute
\newcommand{\mydate}{\ \today \ - \number\hour :\number\minute}

%
%
\title{Restless pions: orbifold boundary conditions\\
and noise suppression in lattice QCD}
\author{Paulo F.~Bedaque\footnote{{\tt bedaque@umd.edu}}} 
\author{Andr\'{e} Walker-Loud\footnote{{\tt walkloud@umd.edu}}} 
\affiliation{Maryland Center for Fundamental Physics\\
    Department of Physics,
    University of Maryland,
    College Park, MD 20742} 

\date{\mydate}

\preprint{UMD-40762-395}
\begin{abstract}
The study of one or more baryons in lattice QCD is severely hindered by the exponential decay in time of the  signal-to-noise ratio.  The rate at which the signal-to-noise  decreases is a function of the  the pion mass. More precisely, it depends on the minimum allowed pion energy in the box, which, for periodic boundary conditions, is equal to its mass.  We propose a set of boundary conditions, given by a ``parity orbifold'' construction, which eliminates the zero momentum pion modes, raising the minimum pion energy without altering the QCD ground state, and thereby improving the signal-to-noise ratio of (multi)-baryon correlation functions at long Euclidean times.  We discuss variations of these ``restless pions" boundary conditions and focus on their impact on the study of nuclear forces.
\end{abstract}
\maketitle

%
%
\section{Introduction}
Lattice QCD studies of heavy systems are plagued by large statistical noise.  The signal-to-noise ratio of a correlation function created by an operator with $n$ quark--anti-quark pairs  decreases with (Euclidean) time as $e^{-(E-\frac{n}{2}m_\pi)t}$, where $E$ is the mass of the state under consideration. This is a particularly nasty problem for the recent studies of nucleon-nucleon \cite{nplqcd_NN,aoki_NN}  and hyperon-nucleon \cite{nplqcd_hyperons} forces with lattice QCD.  The large statistical error renders the numerical information at large times useless.  Compounded with this problem, at early times the correlators are contaminated by excited states and so there is only a very narrow range of time slices left containing useful information. In unquenched calculations~\cite{nplqcd_hyperons,nplqcd_NN} the statistical noise allows for semi-quantitative results only,  even after the computation of thousands of fermion propagators. These errors are also much larger than  finite-volume  \cite{bedaque_finiteNN} and finite lattice spacing~\cite{Chen:2007ug} effects in these observables.

We propose here a scheme to alleviate this signal-to-noise problem.  We start with the simple observation  that the statistical noise is dominated by the energy of the lightest pion states.  Periodic boundary conditions allow a pion zero mode and thus the lowest pion energy is equal to its mass.  If one were to impose anti-periodic boundary conditions \textit{for all three pions}, then the pion zero-modes are forbidden and the minimum energy is given by (assuming anti-periodic boundary conditions in all three spatial directions)
\begin{equation*}
	E_\pi = \sqrt{ 3 \left( \frac{\pi}{L} \right)^2 + m_\pi^2}\, ,
\end{equation*}
with $L$ the size of each spatial direction.  Thus it is clear that the signal can be improved by using these ``restless pions" boundary conditions.  There are other applications for which these restless pions are useful.  In addition to the obvious benefit to spectroscopy studies~\cite{spectroscopy}, the advantage of anti-periodic boundary conditions for the pion in the extraction of the $K\rightarrow\pi\pi$ amplitude using the Lellouch--L\"uscher method~\cite{lellouch} were pointed out in references \cite{kim1,kim2}.

In lattice calculations, one does not have direct control over the hadronic boundary conditions. What can be controlled at will are the boundary conditions of the quark and gluon fields.  However, it is not obvious which modifications of the quarks and gluons at the boundary implies an anti-periodic boundary condition for all three of the pions; anti-periodic boundary conditions for the neutral pion have remained elusive.  Obvious choices get tantalizingly close to the desired restless pions but upon close inspection, have undesired consequences.  For instance, twisted boundary conditions~\cite{Bedaque:2004kc}  allow for continuous momentum transfer by providing hadrons a momentum kick at the boundary.  But as we will explain in sec.~\ref{sec:SigToNoise}, these twisted boundary conditions do not effect the signal-to-noise issue we are interested in.  The G-parity boundary condition suggested in references \cite{kim1,kim2} breaks both the spatial subset of the hypercubic rotation invariance as well as chiral symmetry~\cite{Wiese:1991ku}.  
In Ref.~\cite{kim2}, an isospin boundary condition was used, $q(L)=\tau^3 q(0)$, but this leaves the neutral pion unaffected.  This allows for an extraction of the $\Delta\ I=3/2$ $K\rightarrow\pi\pi$ amplitude (in which the pions are in an $I=2$ final state) but it does not serve our purpose of reducing the statistical noise for baryon calculations.  Various ``hybrid boundary conditions" have been employed, first in the numerical study of penta-qaurk states~\cite{Ishii:2004qe}.  In this first implementation, the $u$ and $d$ quarks were given anti-periodic boundary conditions while the $s$ quark was given periodic boundary conditions, allowing for a definitive identification of bound \textit{vs.} scattering states by forcing the $nK$-system to have non-zero relative momentum (the scattering state) while leaving a possible $\Theta^+(uudd\bar{s})$ resonant state unaffected (the bound state).  In a second variation of the hybrid boundary condition, used in the numerical study of charmonium~\cite{Iida:2006mv} and a possible tetra-quark state~\cite{Suganuma:2007uv}, an anti-periodic boundary condition is imposed upon the quarks while the anti-quarks are given periodic boundary conditions.  This allows for the same identification of bound and scattering states for these systems as the first variant, however this second hybrid boundary condition violates charge conjugation invariance.  Furthermore, neither variant of these hybrid boundary conditions will help with the signal to noise issue we want to address.%
\footnote{See section~\ref{sec:SigToNoise} for details.} 
An axial twisted boundary condition, $q(L)=\gamma_5 q(0)$ (and similar choices) provide for anti-periodic pions but additionally make $\sigma\sim\langle \bar q q\rangle$ anti-periodic and alter the QCD pattern of symmetry breaking.

We propose a novel approach to this problem making use of an orbifold boundary condition.  Similar constructions have been employed in the context of the ``chirality problem" in extra-dimensional extensions of the Standard Model~\cite{chirality_problem}, domain-wall fermions~\cite{kaplan,luscher_lectures} and the Schr\"odinger functional formalism~\cite{schroedinger}.  Instead of relating the field values at the two ends of the box ($z=0$ and $z=L$), we impose periodic boundary conditions on an extended box $-L < z < L$.  However, the fields at negative values of $z$ are not independent, but are determined from those with positive $z$.  By appropriately choosing a relation between the quark and gluon fields in these two halves of the lattice (the orbifold condition), we can enforce a $\pi(-z)=-\pi(z)$ condition, eliminating the zero momentum mode for all pions, making them restless.

%
%
\section{Signal-to-noise ratio estimates \label{sec:SigToNoise}}

\bigskip
\begin{figure}[tbp]
\includegraphics[width=0.48\textwidth]{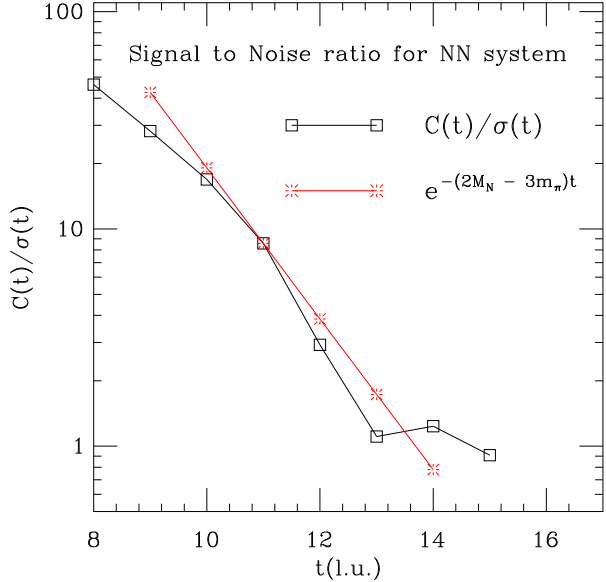}
\noindent
\caption{Log of signal-to-noise ratio of the two-nucleon correlator in the spin singlet channel as a function of Euclidean time (from the calculation described in \cite{nplqcd_N}). The pion mass is about $350$ MeV.  The signal-to-noise estimate of eq.~\eqref{eq:B_signaltonoise} was normalized to the lattice calculation at $t=11$.}
\label{fig:signaltonoise_2}
\end{figure}  

Here we review the  argument estimating the statistical noise for lattice QCD correlation functions~\cite{lepage_error}.  Consider first a nucleon correlator
\beq
C(t)= \langle q(t)q(t)q(t)\ \bar q(0)\bar q(0)\bar q(0)\rangle,\nn\\
\eeq 
where, for clarity, we have suppressed the Dirac, flavor and color indices.  At large times, $C(t)$ is dominated by the intermediate state of lowest energy with the quantum numbers of the nucleon:
\beq
C(t)\  \stackrel{\small t\rightarrow\infty}{\longrightarrow}\  A e^{-Mt},
\eeq
where $M$ is the nucleon mass. In a Monte Carlo calculations, $C(t)$ is estimated by an average over $N$ gauge configurations
\bea
   C(t) \cong\bar C(t)&=& \frac{1}{N}\sum_A S_A(t)S_A(t)S_A(t)\nn\\
                      &\equiv& \langle S_A^3(t) \rangle,
\eea 
where $S_A(t)$ is the quark propagator in each one of the gauge configurations, $A$. The variance in this estimate is given by
\bea
\sigma_C^2(t) &=& \frac{1}{N}\sum_A|S_A(t)S_A(t)S_A(t)-\bar C(t)|^2 \nn\\
	&=& \langle S_A^3(t)S_A^{\dagger\ 3}(t)\rangle - |\bar C(t)|^2.
\eea 
For large times, $\bar C^2(t) \sim e^{-2Mt}$, while the large time behavior of $\langle S_A^3(t)S_A^{\dagger\ 3}(t)\rangle$ can be found by noticing that 
\bea\label{eq:S6}
	\langle S_A^3(t)S_A^{\dagger\ 3}(t)\rangle 
		&=& \langle q^3(t)\bar Q^3(t)\ \bar q^3(0) Q^3(0)\rangle
		\nn\\
		&\stackrel{\small t\rightarrow\infty}{\longrightarrow} & B e^{-3 m_\pi t}\, ,
\eea 
where $Q$ is a fictitious quark with identical quantum numbers and properties of the $q$ quarks.%
\footnote{This explains why the twisted and hybrid boundary conditions do not help the signal-to-noise problem.  The fictitious $Q$ quarks have the same boundary conditions as the $q$ quarks, and thus the $q\bar{Q}$ and $Q\bar{q}$ mesons have periodic boundary conditions and are allowed a zero-momentum mode.} 
The long time behavior of the correlator in eq.~(\ref{eq:S6}) is then dominated by the intermediate state with the lowest energy with the quantum numbers of three $q\bar{Q}$ mesons.  Since they have the same mass as the $q\bar{q}$ mesons, this lowest energy state is given by three times the pion mass.  Thus, for sufficiently light pions, $\langle S_A^3(t)S_A^{\dagger\ 3}(t)\rangle$ decays at a rate smaller than $C^2(t)$.  The signal-to-noise ratio of the nucleon correlator is then given by
\beq\label{eq:B_signaltonoise}
	\frac{C(t)}{\sqrt{\frac{1}{N}\sigma_C^2(t)}}\ \stackrel{\small t\rightarrow\infty}{\longrightarrow}\ 
		A\sqrt{N} \frac{e^{-Mt}}{e^{-\frac{3}{2}m_\pi t}} 
		\sim \sqrt{N} e^{-(M-\frac{3}{2}m_\pi) t}\, .
\eeq 
We show in fig.~(\ref{fig:signaltonoise_2}) the signal-to-noise ratio in an actual lattice QCD calculation (details of the simulation can be found in Refs.~\cite{nplqcd_N,nplqcd_NN}) as well as the estimate in eq.~(\ref{eq:B_signaltonoise}).

The estimate in eq.~\eqref{eq:B_signaltonoise} is easily generalized for correlation functions of multi-baryons and baryons with strange quarks.  In the case of two-nucleon correlators, for example, the signal-to-noise ratio is proportional to $\sqrt{N} e^{-(2M-3m_\pi)t}$.  Recent lattice studies of nuclear forces (and hyperon-nucleon interactions) were severely hindered by the fast decrease of the signal-to-noise ratio with time \cite{nplqcd_NN,nplqcd_hyperons}. The correlators at short times cannot be used for fitting purposes since it is contaminated by excited states%
\footnote{In the single baryon sector, a significant improvement in the isolation of the ground state and excited states at early times has been achieved with the use of multiple operators combined with quark and gluon smearing~\cite{spectroscopy}.  The equivalent study for operators coupling to multi-nucleon states has not been performed and is anticipated to be significantly more challenging and costly given the larger number of operators and quark contractions.} 
while at later times the statistical noise overwhelms the signal, leaving only a very narrow plateau from which the physics is extracted.  This is in stark contrast to lattice calculations of $\pi\pi$ interactions~\cite{Sharpe:1992pp} and other two-meson systems~\cite{Miao:2004gy}.

%
%
\section{Parity orbifolds}

\begin{figure}[tbp]
  \centerline{{\epsfxsize=2.0in \epsfbox{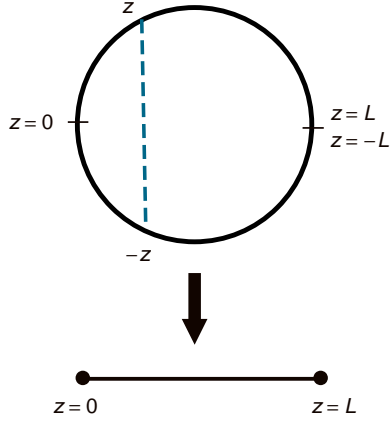}}}
\noindent
\caption{Identification of $z$ and $-z$ points reduces the circle to a line segment.}
\label{fig:orbifold}
\end{figure}

Let us now describe the basic idea of the orbifold construction in the case when only one dimension is orbifolded.
Consider a lattice whose $z$ coordinate belongs to the interval  $[0, L ]$. Extend it to $[-L, L ]$ and identify the points $z=-L$ and $z=L$, effectively turning the interval $[-L, L ]$ into a circle.  Let all fields, 
$\phi(z)$,  satisfy the periodic condition $\phi(L)=\phi(-L)$.  Then identify the points $z$ and $-z$ by relating $\phi(z)$ to $\phi(-z)$, effectively transforming the circle into a line segment (including the boundary) as shown in fig.~(\ref{fig:orbifold}). In the simplest case, $\phi(z)=\pm\phi(-z)$. If the plus sign is chosen, 
$\phi(z)$ will be a linear combination of spatially symmetric wavefunctions,
\beq
	\phi_{+}(z) = \sum_{n=0}^{\infty} A_+^{(n)} \cos \left( \frac{n \pi z}{L} \right)\, .
\eeq 
If, however, the minus sign is chosen 
then the $\phi(z)$ will be a linear combination of anti-symmetric wavefunctions,
\beq
	\phi_{-}(z) = \sum_{n=1}^{\infty} A_-^{(n)} \sin \left( \frac{n \pi z}{L} \right)\, ,
\eeq 
and consequently there is no zero mode for this field. The lowest momentum allowed is $k_{min}=\frac{\pi }{L}$ with an energy of $\sqrt{(\pi / L)^2+m^2}$. This upward shift in the minimum allowed energy value is the desired result.  In order to eliminate the pions at rest we will require that $\mathbf{\pi}(z) =-\mathbf{\pi}(-z)$. The ways to achieve this by imposing orbifold conditions on the quark and gluon fields and the generalization to higher dimensions will be discussed next.
\bigskip

%
%
\subsection{One-dimensional $S^1/\mathbb{Z}_2$ parity orbifold}
In the simplest version of our proposal the orbifold trick is used in only one of the spatial directions. Consider QCD fields in the periodic box $[0,L]\times[0,L]\times[-L,L]\times[0,\beta]$ satisfying the ``parity orbifolding'' condition (the issues we discuss here belong to the infrared regime and we use a continuum notation) 
\bea\label{eq:orbifold1}
A_\mu(t,x,y,z) &=& A_\mu(t,x,y,-z),\ \ {\rm for}\ \mu\neq 3\nn\\
A_3(t,x,y,z) &=& -A_3(t,x,y,-z),\nn\\
q(t,x,y,z) &=& \mathcal{P}_z q(t,x,y,-z),\nn\\
\bar q(t,x,y,z) &=&  \bar q(t,x,y,-z)\mathcal{P}_z\, ,
\eea  
where $\mathcal{P}_z=i\gamma_5\gamma_3$ is the $z$-parity operator corresponding to a reversal of the $z$ direction and we work in Euclidean space.%
\footnote{We use the conventions $\gamma_5^2=\gamma_\mu^2=1$, $\gamma_\mu^\dagger=\gamma_\mu$.} 
The $z$-parity operator $\mathcal{P}_z$ is obtained from the usual parity operator $\gamma_0$, corresponding to a simultaneous reversal of all three spatial axes, combined with a rotation by $\pi$ around the $z$-axis.  The conditions in eq.~\eqref{eq:orbifold1} relate the QCD fields in one side of the box to their parity conjugates in the opposite side. Notice that, since parity is a symmetry of the theory, the contribution to the action from the $z<0$ region is exactly the same as the $z>0$ region and the computational cost of using  the extended box, $[-L,L]$ is the same as that of the smaller box, $[0,L]$. The only effect of the orbifold condition is on the link connecting the $z<0$ and $z>0$ regions. In other words, it acts as  a boundary condition at $z=0$. In fact, consider the orbifolded action in the case of Wilson quarks
\bea\label{eq:lattice_orbifold}
S &=&\kappa \left[\bar q_{-1} (\gamma_3-r) q_1 - \bar q_{1} (\gamma_3+r) q_{-1} \right] +a^4 (\bar q_1 q_1+\bar q_{-1} q_{-1})+\cdots\nn\\
&=& -2 \kappa \bar q_{1} (\gamma_3+r) \mathcal{P}_z q_1 +2 a^4 \bar q_1 q_1+\cdots,
\eea 
where $\kappa$ is the hopping parameter, the index on the quark fields denotes the position in $z$ (the remaining coordinates are implicit) and the dots denote the contributions from the two sides of the bulk, $z>0$ and $z<0$ (which are equal to each other). We see then that the orbifolded $[-L,L]$ lattice is equivalent to a $[0,L]$ lattice with some extra terms residing at the boundary, as is the case with any  boundary condition.%
\footnote{Notice that, contrary to the continuum case, the boundary conditions in lattice field theory are already contained in the action. Different lattice action terms localized at the boundary imply different boundary conditions in the continuum and the relation between them is, in general, a complicated dynamical question.}

Notice that we  could have equally used the opposite $z$-parity operator, $-\mathcal{P}_z$, implementing a reversal of all three spatial axis followed by a rotation by $-\pi$ about the $z$-axis. The difference between rotating by $\pi$ in the positive or in the negative direction amounts to a $2\pi$ rotation which, for spin-1/2 fermions, leads to a minus sign difference between $\mathcal{P}_z$ and $-\mathcal{P}_z$. Physical observables, being quark bilinears, generally do not depend on this sign. As can be seen  in eq.~(\ref{eq:lattice_orbifold}), however, the boundary terms are
linear in $\mathcal{P}_z$ and are able to distinguish between the choice in sign of $\mathcal{P}_z$. This shows that the orbifold condition breaks the $z\rightarrow -z$ symmetry.

The parity orbifold condition on the quark and gluon fields implies orbifold conditions for the hadronic fields. If we identify the pion field with the $\mathbf{\pi} \sim \bar q \gamma_5 \mathbf{\tau} q$ interpolating field, we see that it satisfies the desired
\beq\label{eq:orbifold1_pion}
\mathbf{\pi}(t,x,y,z) = - \mathbf{\pi}(t,x,y,-z)
\eeq 
orbifold condition. In fact, the same condition will follow  if any other pion interpolating field is used like, for instance, $\pi\sim \bar q \mathbf{\tau}q F_{\mu\nu}\tilde F_{\mu\nu}$, since it depends only on the fact that the pion has negative intrinsic parity. In fact, all parity odd operators will satisfy a condition similar to eq.~(\ref{eq:orbifold1_pion}) while the parity even operators will satisfy the analogue equation without the minus sign.  
In particular, the $\sigma$ field $\sigma\sim \langle \bar q q\rangle$ has a zero mode and the QCD pattern of symmetry breaking is not affected by the orbifolding procedure. The nucleon fields satisfy 
\bea\label{eq:orbifold1_N}
N(t,x,y,z) &=& -\mathcal{P}_z N(t,x,y,-z),\nn\\
\bar N(t,x,y,z) &=&  -\bar N(t,x,y,-z)\mathcal{P}_z,
\eea 
as can be seen using the interpolating field $N \sim q q^T\tau_2  C\gamma_5 q$. In the non-relativistic domain, $\mathcal{P}_z = i\gamma_5\gamma_3$ reduces to $\sigma_3$ and the allowed modes for the nucleon are
\beq
N(x,y,z) = e^{i\frac{n_x\pi x}{2L}x+i\frac{n_y\pi y}{2L}y}
\left\{
\begin{array}{cl}
\cos(\frac{n_z\pi z}{L}) \left( \begin{array}{c}1\\0\end{array}\right),
& \ \  n_x, n_y, n_z=0,1,\cdots \\ \\
\sin(\frac{n_z\pi z}{L}) \left( \begin{array}{c}0\\1\end{array}\right)	,
& \ \  n_x, n_y=0,1,\cdots,  n_z=1,2,\cdots .  \end{array}\right.
\eeq  
Notice that only spin up nucleons can be at rest. Consequently we can construct a spin triplet two-nucleon state, like the deuteron, with zero momentum but a spin singlet  two-nucleon state will necessarily have a minimum momentum equal to $\pi/L$.  This asymmetry between spin up and down is a consequence of the breaking of the $z\rightarrow -z$ symmetry discussed above.

\begin{figure}[tbp]
  \begin{center}
    \parbox{2in}{\epsfysize=1.in\epsfbox{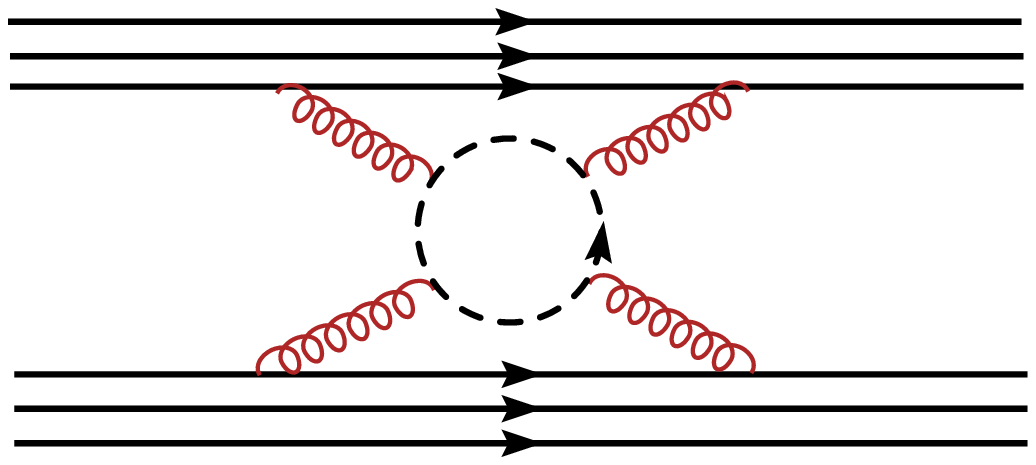}}
    \hspace{0.6in}
    \parbox{2in}{\epsfysize=1.in\epsfbox{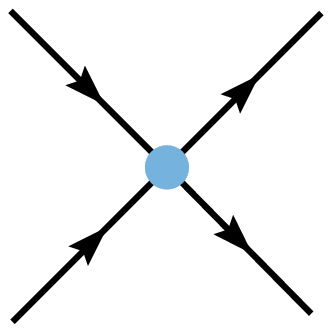}}\\
    \vspace{0.2in}
    \parbox{2in}{\epsfysize=1.05 in\epsfbox{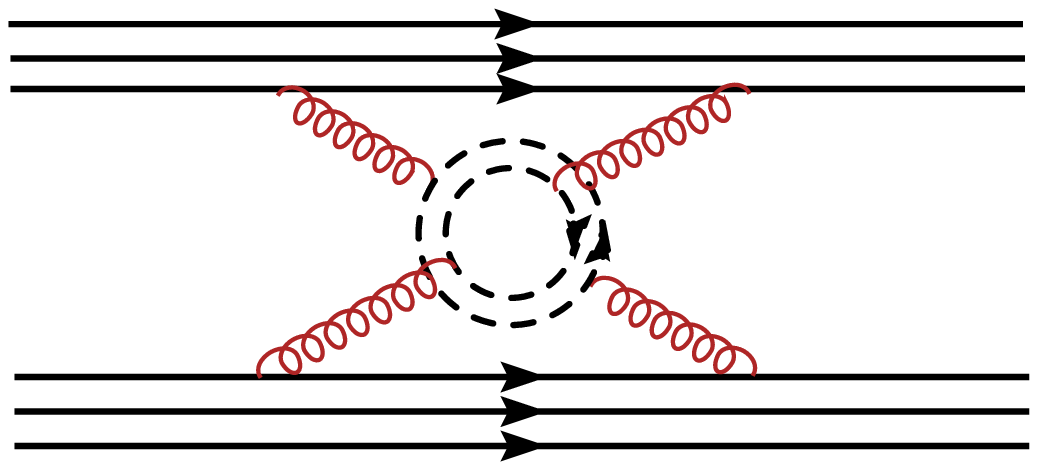}}
    \hspace{0.6in}
    \parbox{2in}{\epsfysize=0.8in\epsfbox{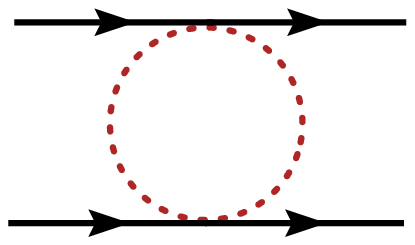}}\\
    \vspace{0.2in}
    \parbox{2in}{\epsfysize=1.in\epsfbox{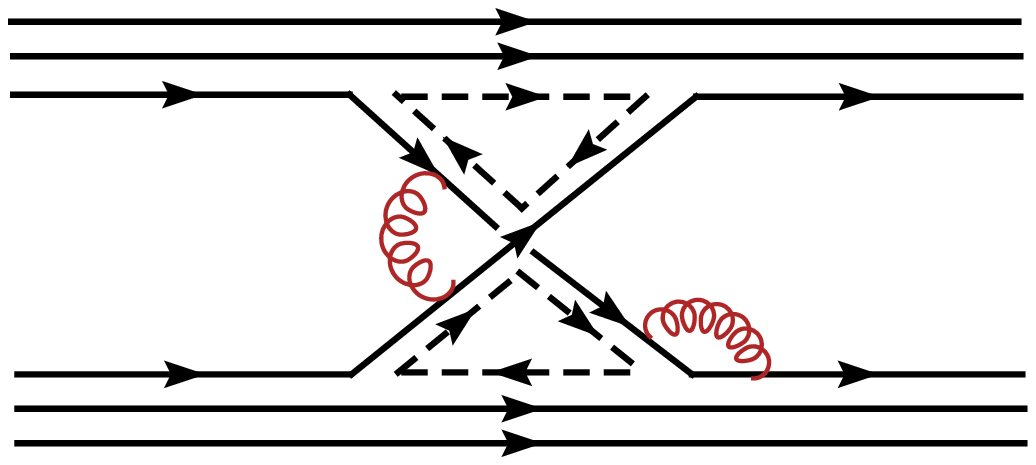}}
    \hspace{0.6in}
    \parbox{2in}{\epsfysize=0.7 in\epsfbox{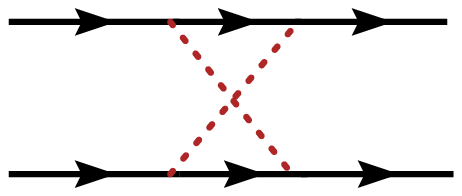}}\\
    \vspace{0.2in}
    \parbox{2in}{\epsfysize=1.in\epsfbox{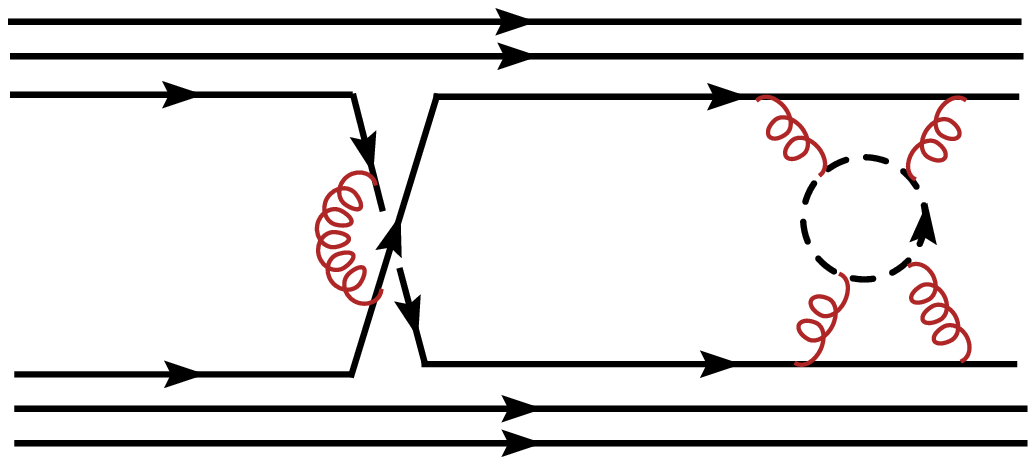}}   
    \hspace{0.6in}
    \parbox{2in}{\epsfysize=0.8in\epsfbox{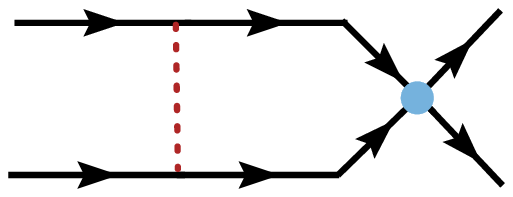}}\\    
  \end{center}
  \caption{\label{fig:partial_orbifold} Examples of two-nucleon graphs containing sea quarks. The left column shows the graphs at QCD level (dotted lines represent sea quarks) and the right column represents the same graphs at the low energy effective theory level. The graphs on the first row are proportional to $e^{-\Lambda_{QCD} L}$, the second and third are proportional to $e^{-m_\pi L}$. The last row shows a graph with a power law dependence on the volume.}
\end{figure}

Unfortunately, the boundary term shown in eq.~(\ref{eq:lattice_orbifold}) is not $\gamma_5$-Hermitian and the fermion determinant is not positive definite. This makes simulations with dynamical quarks satisfying the parity orbifold condition impractical.
However, this method is perfectly suited to implementation in the valence sector only, \textit{i.e.} only on the propagators generated in the background of dynamical configurations.  In refs.~\cite{sachrajda,bedaque_jiunnwei}, it was argued that up to exponentially suppressed corrections, for many channels of interest including baryon-baryon channels, different boundary conditions can be used in the valence and sea sectors of the theory, known as ``partially twisted boundary conditions".  Therefore, gauge configurations generated with sea quarks satisfying periodic boundary conditions can be used with valence quarks satisfying ``parity orbifold" boundary conditions.  Intuitively, the possibility of using  different boundary conditions for sea and valence quarks  follows from the observation that sea quarks can ``notice'' their different boundary conditions only if they propagate around the lattice. But, for observables without annihilation diagrams, the propagation of sea quarks around the lattice is  suppressed by $e^{-m L}$, where $m$ is the mass of the lightest hadron made of sea quarks or a mixture of valence and sea quarks.  In our case, this is the pion mass.  This argument is better appreciated by looking at the graphs in fig.~(\ref{fig:partial_orbifold}), which display examples of processes contributing to baryon-baryon scattering. Only diagrams containing a baryon-baryon intermediate state give rise to power law volume dependence (below the inelastic threshold).  These two intermediate baryons are made of valence quarks and therefore satisfy the orbifold boundary condition.  We stress that the rate at which the signal-to-noise decreases is set by the {\it valence} nucleon and pion masses.

The increase on the pion minimum energy has an additional benefit. With the exception of the relation between two-particle energy levels and the S-matrix, described by the L\"uscher formula, finite volume effects are suppressed by factors of $e^{-E_\pi L}$. An increase on the value of $E_\pi$ is then clearly beneficial. This is specially important for the exponentially supressed correction to the L\"uscher formula where the suppression factor, formally of order $e^{-E_\pi L}$, can be sizable for realistic lattices and periodic pions with $E_\pi=m_\pi$ \cite{bedaque_finiteNN}. These finite volume corrections can be estimated using an extension of chiral perturbation theory adapted to the case where valence and sea quarks obey different boundary conditions in the molds of~\cite{sachrajda,bedaque_jiunnwei,Tiburzi:2005hg,pqchipt,pqqcd_baryons}.

%
%

%
%
\subsection{Three-dimensional parity $T^3/\mathbb{Z}_2$ orbifold }
The method of the previous section can be generalized in order to remove the zero-momentum modes of the pions in all three directions, further improving the signal-to-noise ratio.  The simplest generalization of eq.~(\ref{eq:orbifold1}) is
\bea\label{eq:orbifold2}
A_0(t,\mathbf{r}) &=& A_0(t,-\mathbf{r}),\nn\\
A_i(t,\mathbf{r}) &=& -A_i(t,\mathbf{r}),\ \ {\rm for}\ i=1,2,3\nn\\
q(t,) &=& \mathcal{P} q(t,-\mathbf{r}),\nn\\
\bar q(t,\mathbf{r}) &=&  \bar q(t,-\mathbf{r})\mathcal{P},
\eea 
where $\mathcal{P}=\gamma_0$ is the usual parity operator corresponding to the reversal of all three space directions. While the boundary conditions in eq.~(\ref{eq:orbifold1}) can be seen as a mirror placed at $z=0$, the conditions in eq.~(\ref{eq:orbifold2}) can be visualized as a pin hole located $x=y=z=0$ with a lattice $[-L/2,L/2] \times [-L/2,L/2] \times [-L,L] \times [0,\beta] $.  Again, all three pions obey the odd orbifold condition $\pi(t,\mathbf{r})=-\pi(t,-\mathbf{r})$ but now their minimum energy is $\sqrt{3(\frac{\pi}{L})^2+m_\pi^2}$. Nucleons obey the same conditions as the quarks, $N(t,\mathbf{r})=\gamma_0 N(t,-\mathbf{r})$. Since in the non-relativistic limit  $\gamma_0$ reduces to  $1$, non-relativistic nucleons satisfy periodic boundary conditions and contain zero modes. This property is  very convenient when extracting low-energy phase shifts on the lattice, as with the 3--D parity-orbifolding, there is no restriction on the spin--isospin channels one can study in the ground state and the standard L\"uscher formula relating energy levels to phase shifts is unchanged. 
As it will be exemplified below, the increase in the signal-to-noise ratio is dramatic.

%
%
\section{Impact on lattice calculations}

\subsection{Nuclear force studies}

\bigskip
\begin{figure}[tbp]
\begin{tabular}{cc}
\includegraphics[width=0.48\textwidth]{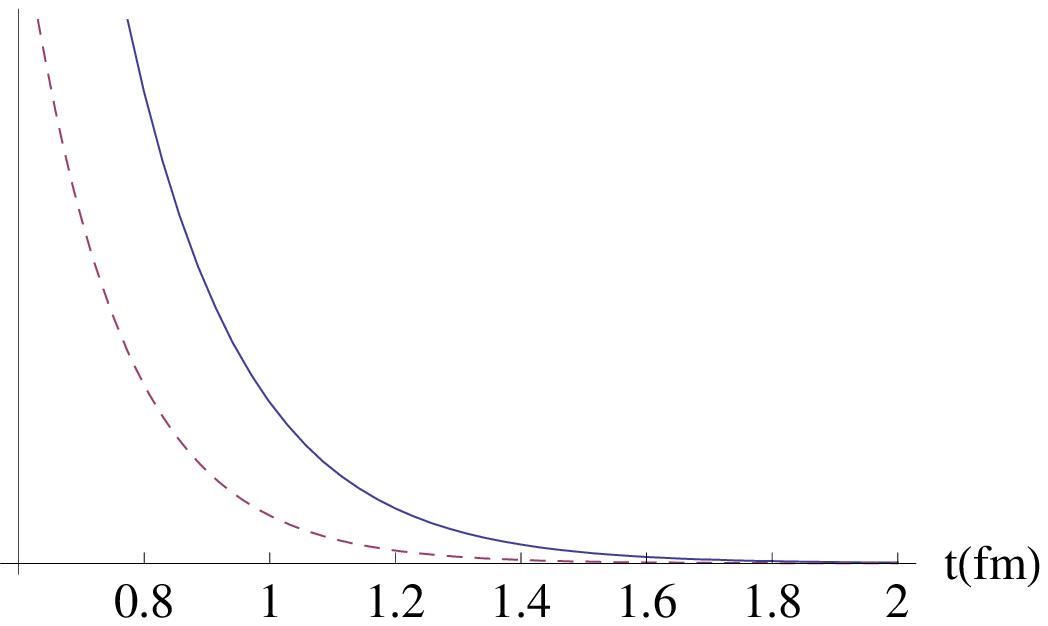}
& \includegraphics[width=0.48\textwidth]{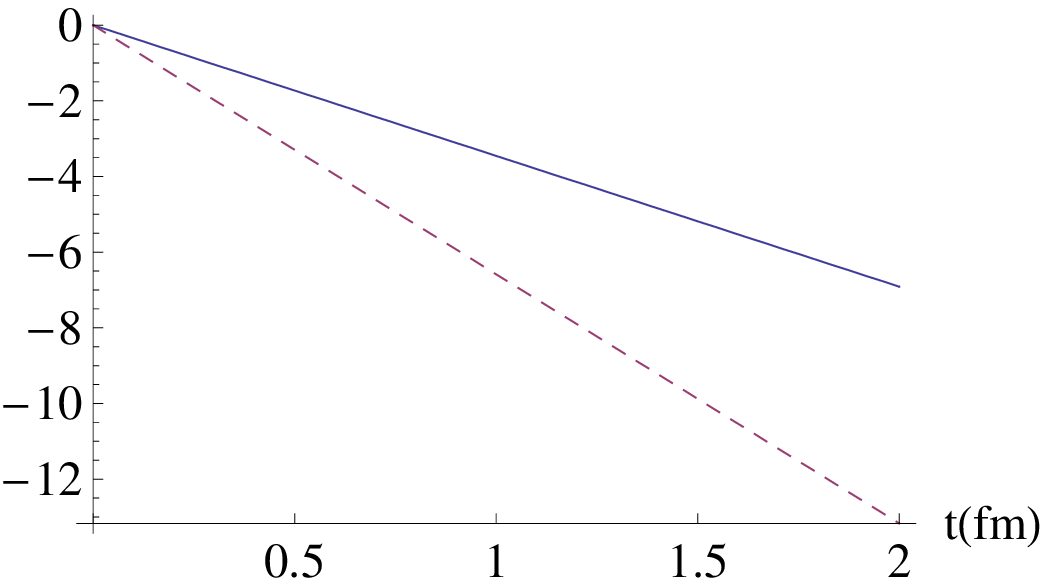}\\
(a) & (b)
\end{tabular}
\noindent
\caption{ Left: Estimate of the signal-to-noise ratio with the $S^3/\mathbb{Z}_2$ orbifold condition
 $e^{-(2M-3\sqrt{(\frac{\pi}{L})^2+m^2})t}$ (solid line) and with periodic boundary conditions $e^{-(2M-3m)t}$ (dashed line) as a function of $t$.
Right: {\it Log } plot of the signal-to-noise ratio with the $T^3/\mathbb{Z}_2$ orbifold condition
 $e^{-(2M-3\sqrt{3(\frac{\pi}{L})^2+m^2})t}$ (solid line) and with periodic boundary conditions $e^{-(2M-3m)t}$ (dashed line) as a function of $t$. In both figures the pion mass is $350$ MeV and the box size is $L=2.5$ fm. 
 }
\label{fig:signaltonoise}
\end{figure}  

In order to provide an explicit example, we use the values of the parameters used in  \cite{nplqcd_NN} to estimate the  impact of the method advocated here in   the expected rate with which the signal-to-noise ratio decreases with increasing time. We disregard the interaction energy between the hadrons and approximate the energy of the two-nucleon state by $\approx 2M$. The energy of the three-pions is approximated  by $\approx 3m_\pi$ when periodic boundary conditions are used, $3\sqrt{(\frac{\pi}{L})^2+m_\pi^2}$ if the $S^1/\mathbb{Z}_2$ orbifold is used and $3\sqrt{3(\frac{\pi}{L})^2+m_\pi^2}$ if the $T^3/\mathbb{Z}_2$ orbifold is used. The result is plotted in fig.~(\ref{fig:signaltonoise}).
  The inclusion of the interaction energy between the two nucleons would change the figure by very little. In fact, for pion masses above $350$ MeV the energy shifts found \cite{nplqcd_NN} are of order of $10-20$ MeV. It is expected, however, that in a narrow band close to the physical value of $m_\pi$ the energy shift should be larger \cite{nplqcd_foundation}, corresponding to the diverging scattering lengths, but still much smaller than the rest mass of the nucleons. Even the modest increase in the pion minimum energy found in the one-dimensional orbifolding has a potential significant impact by noise limited measurements. In the case of the three-dimensional orbifolding that potential improvement is enormous (notice the log scale in the corresponding graph).

\subsection{Impact on $K\rightarrow \pi\pi$}

As pointed out in \cite{kim1,kim2}, the extraction of the $K\rightarrow \pi\pi$ amplitude with the Lellouch-L\"{u}scher method~\cite{lellouch} can benefit from eliminating  pion zero modes. The method to eliminate pions at rest discussed here can only be applied to the $I=2$ channel. In the $I=0$ channel, the use of different boundary conditions in the valence and sea sectors alters the amplitude by factors that are {\it not} exponentially suppressed. Of course, a modified chiral perturbation theory taking into account the differences of the valence and sea sectors can still be used to relate the results of such a lattice calculation with the real world QCD amplitude.

%
%
\section{Discussion}

We have introduced ``restless pions" boundary conditions designed to reduce the rapid degradation of the signal-to-noise ratio which plagues studies of heavy systems with lattice QCD.  We have shown how these boundary conditions can be implemented with a parity-orbifold construction in either one or three spatial dimensions.  Unfortunately, the action at the boundary is not $\gamma_5$-Hermitian and so this particular construction is not suitable for the sea sector.  However, this method is perfectly suited for implementation of the valence fermions.  For non-scalar channels, the difference in sea and valence boundary conditions is felt only by exponentially small terms.  The numerical cost of implementing these parity-orbifolded valence propagators is the same for propagators with (anti)-periodic boundary conditions, as the fields in each half of the bulk are not independent, and therefore the implementation is achieved with a special boundary condition on the non-doubled lattice.

%
%
\begin{acknowledgments}
We would like to thank T. Cohen and K. Orginos for conversations on this subject and the NPLQCD collaboration for the use of their data in fig.~(\ref{fig:signaltonoise_2}).  This research was supported in part by the U.S. Dept. of Energy under grant no. DE-FG02-93Er-40762.
\end{acknowledgments}

  
\end{document}